\documentclass[aps,floats,prl,superscriptaddress,
twocolumn]{revtex4-1}

\usepackage{amsfonts,amsmath} \usepackage{bm} \usepackage{dcolumn}
\usepackage{epsfig} \usepackage{latexsym}

\usepackage[dvipsnames]{xcolor}

\def\be{\begin{equation}}
\def\ee{\end{equation}}
\def\bea{\begin{eqnarray}}
\def\eea{\end{eqnarray}}
\def\bp{\begin{pmatrix}}
\def\ep{\end{pmatrix}}

\begin{document}

\title{Lee-Yang paradigm of phase transition in eigenstate thermalized systems}

\author{Yongjiang Xu}
\affiliation{School of Quantum \& Kavli Institute of Theoretical Sciences, University of Chinese Academy of Sciences, Beijing 100190, China}

\author{Weixin Sun}
\affiliation{School of Quantum \& Kavli Institute of Theoretical Sciences, University of Chinese Academy of Sciences, Beijing 100190, China}

\author{Chushun Tian}
\affiliation{Institute of Theoretical Physics, Chinese Academy of Sciences, Beijing 100190, China}

\author{Huajia Wang}
\email{wanghuajia@ucas.ac.cn}
\affiliation{School of Quantum \& Kavli Institute of Theoretical Sciences, University of Chinese Academy of Sciences, Beijing 100190, China}
\affiliation{Peng Huanwu Center for Fundamental Theory, Hefei, Anhui 230026, China}

\begin{abstract}
As phase transitions in isolated quantum systems remain elusive, here we show how a thermodynamic-like phase transition, falling into the Lee-Yang paradigm,
 can arise in systems displaying eigenstate thermalization. Specifically, we show that in holographic conformal field theories, the eigenstate expectation of the auto-correlation function
 can be mapped to the partition function ${\cal Z}_{{gauge}}(z)$ of a virtual interacting instanton gas, with the conformal mapping of the imaginary time: $z=1-e^{-\tau}$ and the central charge $c$ mimicking the instanton fugacity and volume,
  respectively. We find that akin to the Lee-Yang paradigm, for $c\to\infty$ a pair of complex conjugate zeros of ${\cal Z}_{{gauge}}(z)$ move to the real axis located at the famous forbidden singularity. Passing through the singularity the system transits from the low- to high-fugacity phase, accompanied by dramatic changes in scaling behaviors of the free energy and dominant microscopic configurations. Our findings indicate that physics of phase transitions from eigenstate thermalization is very rich.
\end{abstract}

\date{\today}

\maketitle

Phase transition is a vehicle advancing the development of many-body physics, and finds considerable practical applications \cite{Chaikin95}. Despite a matter carries a huge amount of microscopic degrees of freedom and thereby exhibits extreme complexity, macroscopically it can be characterized by only a few observables. In the thermodynamic limit, when non-analyticity or singularity of those observables develops at some critical parameter, the system's properties undergo profound changes and a phase transition occurs. Canonical theories of phase transitions are built upon statistical ensembles, which though useful are of fictitious nature. In particular, they do not apply to isolated systems ubiquitous in Nature and experiments, which are described instead by pure states in quantum mechanics. Although studies of statistical physics of isolated systems were initiated almost a century ago \cite{von Neumann29}, not until recent years have substantial theoretical \cite{Popescu06,Lebowitz06,Deutsch91,Srednicki94,Rigol08,Rigol16,Borgonovi16} and experimental \cite{Kaufman16} progresses been achieved. So far, the focus has been the general aspects of the emergence of a thermodynamic description from a pure state.

A major step forward is to explore phase transition phenomena intrinsic to the isolated nature, whose physics might be rich. In fact, when restricted to the ground state, isolated systems already exhibit the well-known quantum phase transition \cite{Sachdev99}, manifesting in profound changes in the ground state properties. While new statistical principles aiming at isolated quantum systems -- notably the eigenstate thermalization hypothesis (ETH) \cite{Deutsch91,Srednicki94,Rigol08}, suggest that thermal equilibrium phenomena can emerge from excited eigenstates, it is natural to expect that in an eigenstate thermalized system, even richer phase transition phenomena may arise from the interplay between the quantum and emergent thermal effects. This notwithstanding, to address this issue is intellectually challenging. In particular, ETH works directly on the expectation of an observable at an eigenstate. Consequently the partition function, as the building block of various phase transition theories, do not enter the theoretical setup. In addition, within the framework of ETH it remains elusive how the thermodynamic limit is achieved and subsequently leads an observable to displaying a singularity --- a key characteristic of phase transitions.

Phase transition in isolated quantum systems is not only of fundamental interests to statistical mechanics, but also central to some outstanding problems in other fields, e.g. conformal field theories (CFTs), the black hole information paradox \cite{Page:1993wv,Hawking:1975IP,Hawking:1976IP}, etc. Notably, in CFTs, owing to the state-operator correspondence, expectations of observables in energy eigenstates are mapped to correlation functions whose properties are more tractable, allowing in-depth investigations of ETH. Interestingly, when ETH is applicable to eigenstate expectations of auto-correlation functions,
the singularity due to the coincidence of operators develops an infinite number of thermal images in the domain of imaginary time $\tau$. Being in sharp conflict with the unitarity of the isolated state, these additional thermal singularities are dubbed forbidden singularities, and their appearance resembles the celebrated black hole information paradox. These singularities have received extensive studies in 2d holographic CFTs, especially for the auto-correlation of a light (spinless) primary operator $O_{L}$:
\begin{equation}\label{eqn:auto-correlation}
    f_{E}(\tau) = \langle E\rvert O_{L}(0)O_{L}(\tau)\lvert E\rangle
\end{equation}
at the high energy eigenstate $\lvert E\rangle = O_{H}(-\infty)\lvert \Omega\rangle$, prepared by inserting a heavy primary operator $O_{H}$ into the vacuum $\lvert\Omega\rangle$ \cite{Fitzpatrick:2014,Fitzpatrick:2015, Fitzpatrick:2016ive, Fitzpatrick:2016mjq, Chen:2017,Wang:2018,Collier:2018exn}. This eigenstate correlation is believed to capture important aspects of holographic CFTs, with $O_H$ creating a micro-state whose bulk dual is the Banados-Teitelboim-Zanelli black hole, and $O_{L}$ probing the spacetime background excited by that black hole. Notwithstanding that the discovery of the forbidden singularity is motivated by eigenstate thermalization, such singularity is rarely addressed from thermodynamic perspectives. In particular, it is unknown whether it leads to a phase transition.

\begin{figure}[t]
\centering
\includegraphics[width=1\linewidth]{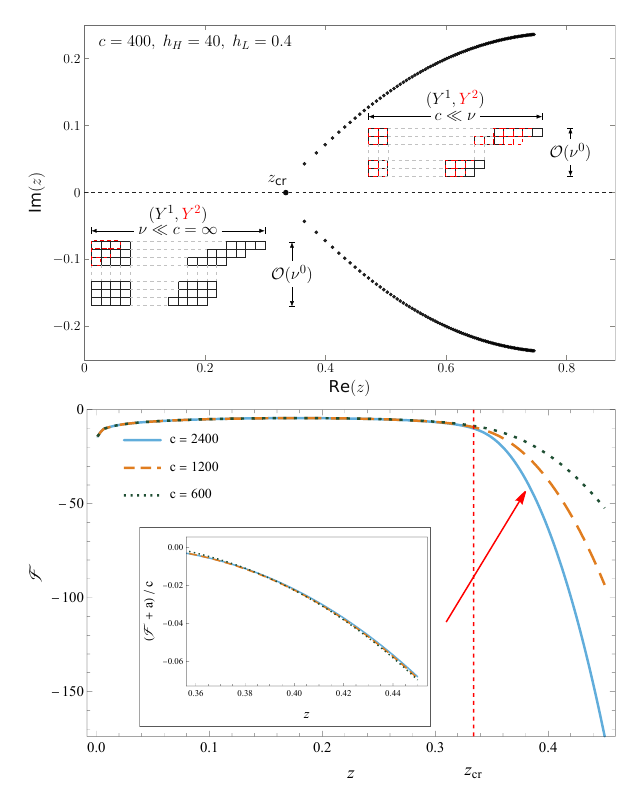}
\caption{Top: numerical experiments confirm that for $c\to\infty$ a pair of complex conjugated zeros of ${\cal Z}_{gauge}(z)$ move into the real axis located at the forbidden singularity $z_{\rm cr}$, separating the low- and high-fugacity phase whose dominant configurations $(Y^1,Y^2)$ differ. (For large but finite $c$ the distance of that zero pair to the real axis is $\propto {1/\sqrt{c}}$ \cite{Wang:2026}.) Bottom: numerical experiments also confirm that in the two phases the free energy ${\cal F}(z)$ has distinct large $c$ behaviors. Inset: for $z>z_{\rm cr}$ numerical data of ${1\over c} {\cal F}(z)$ with distinct $c$ collapse into a single curve. (A small finite-$c$ correction ${a\over c},\,a=0.46$ is found numerically.) }\label{fig:LY}
\end{figure}

In this Letter, we develop systematic analytical treatments and perform numerical experiments to show that the appearance of the forbidden singularity in $f_E(\tau)$ falls into the Lee-Yang paradigm for thermodynamic phase transitions \cite{Yang:1952be,Lee:1952ig} (cf.~Fig.~\ref{fig:LY}). Utilizing the Alday-Gaiotto-Tachikawa (AGT) correspondence \cite{Alday:2009aq}, we trade $f_E(\tau)$ to the partition function $\mathcal{Z}_{gauge}(z)$ of an emergent interacting gas of instantons in some gauge theory, with $z=1-e^{-\tau}$ and $c$  simulating the fugacity and volume, respectively. That gas has a configuration represented by a pair of Young tableaux $(Y^{1},Y^{2})$, and when $c$ is large the instantons may proliferate with number $\nu\gg 1$. Akin to the Lee-Yang paradigm, for $c\to \infty$ effecting the thermodynamic limit, we find that a pair of complex conjugated zeros of $\mathcal{Z}_{gauge}$ moves into the real axis, locating exactly at the forbidden singularity $z_{\rm cr}$. The singularity then effects a critical point, separating two phases with distinct dominant configurations and carrying distinct ``thermodynamic'' properties. In one phase ($z<z_{\rm cr}$), the ``free energy'' ${\cal F}(z)$ approaches a limit for $c\to\infty$, and in dominant configurations the horizontal size of $Y^1$ is large $\sim \nu\,(<c)$, while the size of $Y^2$ is small in both horizontal and vertical directions. In the other phase ($z>z_{\rm cr}$), ${\cal F}(z)\propto c$ and diverges for $c\to\infty$; in the dominant configurations, the horizontal size of both $Y^1$ and $Y^2$ is large $\sim \nu\gg c$.

Now we outline our analytical theory, and refer to the company paper \cite{Wang:2026} for full details. The eigenstate auto-correlation function $f_E(\tau)$ consists of two ``probe'' operators $O_L$ at $x=0$, being separated by imaginary time $\tau$. In CFTs, via the transformation: $w\to 1-e^{iw},\,w=x+i\tau$, $f_E(\tau)$ is conformally equivalent to a 4-point correlation function on the plane. By performing the operator product expansion of $O_H O_H$ and $O_L O_L$, the latter correlation function is decomposed into a sum of Virasoro conformal blocks, which are universal kinematic objects in 2d CFTs:
\bea \label{eq:OPE}
f_E(\tau)&\propto & \langle O_L(0) O_L(z) O_H(1) O_H(\infty)\rangle\nonumber\\
&=& \sum_{I} C_{I}\mathcal{V}_{I}(z) \bar{\mathcal{V}}_I(\bar{z}),
\eea
where $C_I$'s are expansion coefficients. In CFTs when $c$ is large there are many local degrees of freedom. The notion of heavy and light operators is defined with respect to this condition. To be specific, the scaling dimension of $O_H$ is
$h_H=\mathcal{O}(c)$,
and that of $O_L$ is $h_L=\mathcal{O}(1)$. This is called the heavy-light limit. For holographic CFTs $f_E(\tau)$ has highly universal properties in this limit. In particular, the decomposition (\ref{eq:OPE}) is dominated by the vacuum contribution ($I=vac$):
\be\label{eq:vac_dom}
f_E(\tau) \sim \mathcal{V}_{vac}(z)\mathcal{V}_{vac}(\bar{z}).
\ee

In the heavy-light limit important results about $\mathcal{V}_{vac}(z)$ have been obtained \cite{Fitzpatrick:2014,Fitzpatrick:2015, Fitzpatrick:2016ive, Fitzpatrick:2016mjq, Chen:2017,Wang:2018,Collier:2018exn}. In particular, as an evidence for eigenstate thermalization, explicit singularities of $\mathcal{V}_{vac}(z)$ and thereby of (\ref{eq:vac_dom}) were found at
\be
\label{eq:singularities_z}
z_{\rm cr}=1-e^{-\tau_{\rm cr}},\;\;\text{for} \,\,c\to \infty,
\ee
where $\tau_{\rm cr}$ is the inverse temperature emerging from eigenstate $|E\rangle$,
\be\label{eq:singularities}
\tau_{\rm cr} = 2\pi /\alpha_H, \;\;\alpha_H =\sqrt{24h_H/c-1} \propto \sqrt{E/c}.
\ee
So the forbidden singularity stems from the first thermal image as predicted by eigenstate thermalization \cite{note:image}.

\begin{figure}[b]
\includegraphics[width=0.6\linewidth]{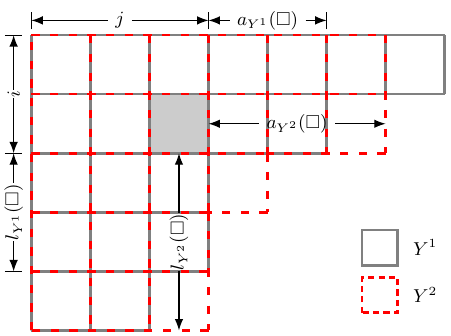}
\caption{Definition of integers $\lbrace i,j,\ell_{Y^{1,2}},a_{Y^{1,2}}\rbrace$ for given $\Box \in Y^1$ (gray). For given $\Box \in Y^2$ these integers are defined in the same way. The values of $\ell_{Y^{1,2}},a_{Y^{1,2}}$ may be negative.}\label{fig:YD}
\end{figure}

Below we show that even though at the eigenstate $|E\rangle$ no quantum coherence is lost, a thermodynamic-like phase transition occurs at $z_{\rm cr}$, and thus Eq.~(\ref{eq:singularities_z}) attains a complete different physical meaning, namely the critical point of the phase transition. To this end we apply the AGT correspondence --  an exact relation between Virasoro blocks and 4d supersymmetric $\text{SU}(2)$ gauge theory partition functions \cite{Alday:2009aq,Nekrasov:2002qd,Nekrasov:2003rj}:
\be\label{eq:AGT}
\mathcal{V}_{vac}(z) = \mathcal{Z}_{gauge}(z)
\ee
holding for generic values of $\lbrace h_H,h_L,c\rbrace$. Thanks to supersymmetry the right-hand side is organized into contributions of gauge theory instantons, whose micro-states are represented by a pair of Young tableaux $(Y^1,Y^2)$. Each Young tableau is represented as $Y^\alpha=\{Y^\alpha_1,Y^\alpha_2,\cdots\}$, where $Y^\alpha_X$ is the length of the $X$-th row, and the size $|Y^\alpha|=\sum_X Y^\alpha_X$. The total size, $|Y^1|+|Y^2|$, gives the instanton number $\nu$. The gauge theory parameters are determined by the CFT parameters \cite{Wang:2026}. Up to an unimportant overall factor,
\bea\label{eq:fugacity_expansion}
\mathcal{Z}_{gauge}(z) &\propto&  \sum^{\infty}_{\nu=0} z^\nu {\cal Z}_\nu(c), \nonumber\\
{\cal Z}_\nu(c)&=&\sum_{|Y^1|+|Y^2|=\nu} e^{-\mathcal{I}\left[Y^1,Y^2\right]}.
\eea
It gives the operator product expansion of the eigenstate auto-correlation. From the CFT side the coefficients ${\cal Z}_\nu$ can be computed by using Zamolodchikov's recursive relation \cite{Zamolodchikov:1984eqp,Zamolodchikov:1987avt}, while from the gauge theory side along which we proceed, they are the sum over contributions from all micro-states of $\nu$ instantons with the action:
\bea\label{eq:Nekrasov}
\mathcal{I}[Y^1,Y^2] = -\sum_{\alpha=1}^2 \sum^4_{\gamma=1}\sum_{\Box\in Y^\alpha}\ln{\left(a_\alpha+\epsilon_1 i+\epsilon_2 j-\mu_\gamma\right)}\quad\nonumber\\
+ \sum_{\alpha,\beta=1}^2 \Big(
\sum_{\Box\in Y^\alpha}\ln{\left(a_\alpha-a_\beta-\ell_{Y^\beta}(\Box)\epsilon_1+(a_{Y^\alpha}(\Box)+1)\epsilon_2\right)}\nonumber\\
+\sum_{\Box\in Y^\beta}\ln{\left(a_\alpha-a_\beta+(\ell_{Y^\alpha}(\Box)+1)\epsilon_1-a_{Y^\beta}(\Box)\epsilon_2\right)}\Big).\qquad
\eea
Here each logarithm is defined on a box $\Box\in Y^{1,2}$. The integers $\ell_{Y^\alpha},a_{Y^\alpha}$ are determined by the box ``coordinates'' $(i,j)$ and the shape of $Y^\alpha$ (Fig.~\ref{fig:YD}). The boxes in $Y^{1,2}$ span a configuration space of $\nu$ instantons. The values of $\lbrace \epsilon_{1,2}, a_{1,2}, \mu_{1,2,3,4}\rbrace$ are gauge theory parameters, which are determined by CFT parameters \cite{Wang:2026}. Hereafter we consider those parameters as given in Table \ref{tab:gauge_HL}. They correspond to the heavy-light limit of the CFT parameters, which is related to but distinct from the so-called Nekrasov-Shatashvili limit \cite{Nekrasov:2009rc}. Physics in the heavy-light limit remains largely unexplored.

\begin{table}[!ht]
\centering
\caption{Parameters in the heavy-light limit
}\label{tab:gauge_HL}
\begin{tabular}{cccccccc}
\hline\hline
$\epsilon_1$ & $\epsilon_2$ & $a_1$ & $a_2$ & $\mu_{1}$ & $\mu_2$ & $\mu_{3}$ & $\mu_4$ \\
\hline
\,\,$c$\,\, & \,\,$6$\,\, & \,\,$c/2$\,\, & \,\,$-c/2$\,\, & \,\,$c/2$\,\, & \,\,$3c/2$\,\,& \,\,$c/2$\,\, & \,\,$\left(1/2+i\alpha_H\right)c$\,\,\\
\hline
\hline
\end{tabular}
\end{table}

In Eq.~(\ref{eq:Nekrasov}), a logarithm in the first line or second/third line ($\alpha=\beta$) depends only on the instanton coordinates in a Young tableau, and thus simulates an external potential exerting on the instanton. Whereas a logarithm in the  second/third line ($\alpha\neq\beta$) is governed by two instantons in distinct Young tableaux, and thus simulates an instanton interaction. As such an interacting instanton gas emerges from the eigenstate $|E\rangle$,  with its grand partition function and fugacity being $\mathcal{Z}_{gauge}(z)$ and $z$, respectively. Despite this analogy the external potential and instanton interaction are exotic, and to obtain their explicit forms is an intractable task.

To proceed we divide the logarithms in Eq.~(\ref{eq:Nekrasov}) into two classes. If a logarithm is defined on $\Box$ satisfying either (a) or (b) below:
\bea
(a)\,\,\Box\equiv (i,j) \in Y^1: \qquad\quad\qquad\qquad\qquad\qquad\qquad\qquad
\nonumber\\
(i-1)\ell_{Y^1}(\Box)\ell_{Y^2}(\Box)(\ell_{Y^2}(\Box)-1)=0; \qquad\qquad\nonumber\\
(b)\,\,\Box\equiv (i,j) \in Y^2: \qquad\quad\qquad\qquad\qquad\qquad\qquad\qquad
\nonumber\\
(i-1)(i-2)\ell_{Y^2}(\Box)(\ell_{Y^1}(\Box)+1)(\ell_{Y^1}(\Box)+2)=0,\,\,\nonumber
\eea
we call it the boundary logarithm, and otherwise call it the bulk logarithm. For the parameters in Table \ref{tab:gauge_HL}, for large $c$ the bulk logarithm is $\sim \ln c$ and the boundary logarithm is ${\cal O}(c^0)$. Then we can show \cite{Wang:2026} the following result, that essentially results from supersymmetry: \\
\\
{\bf Theorem.} {\it Let the CFT parameters be those in the heavy-light limit. Then for any $\nu$ and for any $(Y^1,Y^2)$, all leading $\ln c$ factors in the boundary logarithms cancel out, and the same happens to the bulk logarithms and ${\cal I}[Y^1,Y^2]$.}\\
\\
\noindent It implies that for arbitrary $\nu$ and $(Y^1,Y^2)$, the following limit is well-defined:
\be\label{eq:regime_1}
\lim_{c\to\infty}\mathcal{I}[Y^1,Y^2]=\mathcal{I}_0[Y^1,Y^2],\quad |Y^1|+|Y^2|=\nu.
\ee
which leads to the well-defined limit of: $\lim_{c\to\infty}{\cal Z}_\nu(c)=\sum_{|Y^1|+|Y^2|=\nu} e^{-\mathcal{I}_0[Y^1,Y^2]}$. In Ref.~\cite{Wang:2026} we further show that there exists a constant:
\be
\label{eq:convergence_radius}
r_0=\overline{\lim_{\nu\to\infty}}\,\left|
\lim_{c\to\infty}{\cal Z}_\nu(c)
\right|^{-{1\over\nu}},
\ee
such that for any $z$ satisfying $|z|<r_c$, the infinite series of functions (of $c$ !) $\sum_{\nu=0}^{\infty}z^\nu{\cal Z}_\nu(c)$ uniformly converges in $c$, and as a result the large $c$ limit of $\mathcal{Z}_{gauge}(z)$ exists:
\bea\label{eq:phase_1}
\lim_{c\to\infty}{\cal Z}_{gauge}(z)=\sum_{\nu=0}^{\infty}z^\nu \lim_{c\to\infty}{\cal Z}_\nu(c)\qquad\qquad\nonumber\\
= \sum_{\nu=0}^\infty z^\nu \sum_{|Y^1|+|Y^2|=\nu} e^{-\mathcal{I}_{0}\left[Y^1,Y^2\right]}, \quad |z|<r_c.
\eea
Whereas $\mathcal{Z}_{gauge}(z)$ has no large $c$ limit for $|z|>r_0$. So the large $c$ behaviors of $\mathcal{Z}_{gauge}(z)$ for $z<r_0$ and $z>r_0$ differ significantly. Mathematically, this resembles that in Lee-Yang paradigm for the liquid-gas phase transition, in the thermodynamic limit the analytic structures of the fugacity expansion of the partition function change dramatically, as the fugacity moves along the real axis and passes through the critical value \cite{Yang:1952be}. Here we dub the regime with $z<r_0$ (respectively $z>r_0$) the low-fugacity (respectively high-fugacity) phase.

Next, we show analytically that $r_0=z_{\rm cr}$. In doing so, we will understand properties of the low-fugacity phase and physical implications of $r_0$. For $c\to\infty$, in all bulk logarithms $j\epsilon_2,\,a_{Y^{1,2}}\epsilon_2$ are ${\cal O}(c^0)$ and can be ignored. This yields substantial simplifications. In particular, contributions to $\mathcal{I}_0$ are divided into two classes that are determined by the row lengths $Y_X^{1,2}$ completely \cite{Wang:2026}. In one class (A), each contribution depends on the upper boundary size only, taking the form: $\ln \Gamma(Y^\alpha_1+{\cal O}(\nu^0))+2\delta_{2\alpha}\ln \Gamma(Y_0^2+{\cal O}(\nu^0))$; in the other class (B), each contribution depends on the shape of the zigzag boundary of $Y^{1,2}$, taking the form:
$
\ln \Gamma(Y^\alpha_X-Y^\beta_{X+1}+{\cal O}(\nu^0))$ or $(Y^\alpha_{X}-Y^\alpha_{X+1})G^\alpha_X$, with
\be
\label{eq:G}
G^\alpha_X=\ln\frac{\Gamma(X+2)\Gamma(2-\alpha-i\alpha_H)}{\Gamma(X+1-i\alpha_H)}.
\ee
Hereafter we shift the row index of $Y^2$, i.e. $Y^2_X\rightarrow Y^2_{X-1}$.

The action forms for (B)
are invariant under globally horizontal translation of the zigzag boundary: $Y_X^\alpha\rightarrow Y^\alpha_X+T,\, T\in\mathbb{Z}$.
However, such invariance does not exist in the vertical direction. This strong asymmetry results from the heavy-light limit. It motivates us to rescale $Y^{1,2}$ horizontally:
\be\label{eq:stringy}
Y^\alpha_X = \nu \, y^\alpha_X,\;\;\;\sum_{\alpha,X}  y^{\alpha}_X =1,\;\;\;y^{\alpha}_X\geq 0.
\ee
With its substitution into the action forms for (A,B), simple scaling analysis gives ${\cal I}_0\sim \nu\ln \nu$ for $\nu\gg 1$. However, more sophisticated analysis shows \cite{Wang:2026} that contributions at this order cancel out, and instead ${\cal I}_0\sim\nu$. So we can write the coefficient in Eq.~(\ref{eq:fugacity_expansion}) as a path integral \cite{Wang:2026}:

\bea\label{eq:path_integral}
{\cal Z}_\nu\stackrel{c\to\infty}{=}\int d\lambda \int_D \prod_{X,\alpha} dy^{\alpha}_X \; e^{-\nu (f(y^1,y^2)+\lambda (\sum\limits_{X,\alpha} y^{\alpha}_X-1))}\quad
\eea
for $\nu\gg 1$, with $\lambda$ being a Lagrange multiplier enforcing the normalization in Eq.~(\ref{eq:stringy}). Here
\begin{eqnarray}\label{eq:eff}
f(y^1,y^2) &=& \sum^\infty_{X=-1}\bigg(\sum_{\alpha,\beta} \left[d^{\beta}y^\alpha_X\left(\ln (d^{\beta}y^\alpha_X) + \delta_{\alpha\beta} G^\beta_X\right)\right]\nonumber\\
&&\quad\,\,+i\pi\sum_{\alpha} b^\alpha_X y_X^\alpha\bigg),
\end{eqnarray}
where $d^\beta$ is a linear operator defined as $d^{\beta}y^\alpha_X \equiv y^\alpha_X-y^\beta_{X+1}$, $b^\alpha_X\in 2\mathbb{Z}+1$ are numerical constants, and $y^1_{-1}=y^1_0= y^2_{-1}=0$. The domain of integral $D=\{(y^1,y^2):y^\alpha_{2-\alpha}\geq y^\alpha_{3-\alpha}\geq\cdots\geq y^\alpha_{1+K-\alpha}\geq 0\}\subset \mathbb{R}^{2K}$ and $K\to\infty$ eventually.

To calculate the multi-real variable integral $\int_D$ in Eq.~(\ref{eq:path_integral}), we embed the real manifold $D$ into the complex Euclidean space $\mathbb{C}^{2K}$, and use the theory for analysis of several complex variables \cite{Hormander73}. Because $f$ is holomorphic in the domain $\mathbb{C}^{2K}\setminus \cup_{X,\alpha,\beta}\{d^\beta y^\alpha_X=0,\, y^\alpha_X\in \mathbb{C}\}$,
by the Cauchy-Poincar$\acute{\rm e}$ theorem we can keep the boundary of $D$ fixed and deform $D$ into $\tilde{D}$ without crossing hyperplanes: $d^\beta y^\alpha_X=0$, such that $\int_D=\int_{\tilde D}$ and ${\tilde D}$ passes the saddle point solving the set of equations: $\{{\partial f\over \partial y_X^\alpha}+\lambda=0\}$ \cite{Fedoryuk89}. For $K\to\infty$ the solution is found to be \cite{note_saddle}:

\begin{eqnarray}\label{eq:saddle_1}
\tilde{y}^1_X= \tilde{y}_1^1\sum^{X-1}_{j=0} \binom{i\alpha_H-1}{j} \frac{(-e^{-\lambda})^j}{j+1},\;\;\tilde{y}^2_X = 0.\quad
\end{eqnarray}
From this we obtain
$\tilde{y}^1_{X\to\infty}= {e^\lambda \tilde{y}_1^1\over i\alpha_H} [(1-e^{-\lambda})^{i\alpha_H}-1]$.
Imposing the asymptotic boundary condition $\tilde{y}^1_{X\to\infty}=0$ gives the quantization spectrum $\{\lambda_n\}$ of $\lambda$,
\be
\label{eq:quantization_lambda}
e^{-\lambda_n}=1-e^{-2\pi n/\alpha_H},\, n\in\mathbb{Z}.
\ee

It can be shown that irrespective of the explicit form of the saddle point solution, for any $k$ we have $f(k\tilde{y}^1,k\tilde{y}^2)+k\lambda \sum_{\alpha,X}\tilde{y}^\alpha_X=0$. So, upon inserting Eqs.~(\ref{eq:saddle_1}) and (\ref{eq:quantization_lambda}) back to Eq.~(\ref{eq:path_integral}), we find that ${\cal Z}_\nu(c\to\infty)\sim e^{\nu\lambda_1}$, giving

\be
\label{eq:convergence_radius_1}
r_0=e^{-\lambda_1}=1-e^{-2\pi/\alpha_H}.
\ee
Comparing this with Eq.~(\ref{eq:singularities_z}) we conclude that $r_0=z_{\rm cr}$, i.e. the critical point $r_0$ for the transition from low- to high-fugacity phase coincides with the forbidden singularity $z_{\rm cr}$. Furthermore, in the low-fugacity phase $z<z_{\rm cr}$, at every $z$ the ``free energy'' ${\cal F}(z)\equiv -\ln {\cal V}_{vac}(z)=-\ln \mathcal{Z}_{gauge}(z)$ approaches a finite value for $c\to\infty$.

We further compute $\mathcal{V}_{vac}(z)$ numerically at large but finite $c$ using Zamolodchikov's recursive relation \cite{Zamolodchikov:1984eqp,Zamolodchikov:1987avt}. As shown in Fig.~\ref{fig:LY}, $\mathcal{V}_{vac}(z)$ exhibits a trajectory of zeros on the complex $z$-plane, that becomes more densely packed with increasing $c$. In the limit of $c\to \infty$, the zeros coalesce into two branch-cuts whose branch-points fall onto the real-axis and merge at $z=z_{\rm cr}$. This bears a firm analogy to the Lee-Yang paradigm for phase transitions \cite{Lee:1952ig,Yang:1952be}. In addition, numerical experiments confirm that for $c\to\infty$, ${\cal F}(z<z_{\rm cr})$ is finite.

We turn now to the high-fugacity phase $z>z_{\rm cr}$. In this phase ${\cal Z}_{gauge}(z)$ has no large $c$ limit, and the fugacity expansion (\ref{eq:fugacity_expansion}) is dominated by terms with $\nu \gg c\gg 1$. We focus on this chain of inequalities below. Under such condition we find that the rescaling (\ref{eq:stringy}) remains valid, and the leading order of action ${\cal I}[Y^1,Y^2]$ takes a remarkably simple form:
\be\label{eq:eff_2}
\mathcal{I}[Y^1, Y^2]= \frac{c}{6}\ln\left(\nu\over c\right) n(i\alpha_H-n),
\ee
where $n$ is the sum of the row number of $Y^{1,2}$. In particular, $y^{1,2}$ enter only through sub-leading corrections. The scaling ${\cal I}\sim\ln\nu$ differs from ${\cal I}\sim\nu$ in the low-fugacity phase. Substituting Eq.~(\ref{eq:eff_2}) into Eq.~(\ref{eq:fugacity_expansion}) and using the saddle point method to sum over $n$, we obtain \cite{Wang:2026}
\be
\label{eq:Z_nu_high_fugacity}
{\cal Z}_{\nu} {\sim} \left(\nu\over c\right)^{\frac{c}{24}\alpha_H^2},\quad \nu\gg c\gg 1.
\ee
With its substitution the fugacity expansion gives ${\cal F}\sim c$, which is confirmed numerically (Fig.~\ref{fig:LY}). Moreover, summing up $\nu$ by the saddle point method again, we find that $\mathcal{Z}_{gauge} \sim (1-z)^{-\frac{c}{24}\alpha_H^2}$ for $z\to 1^-$,
in agreement with a result found using CFT crossing relations \cite{Collier:2018exn}. Note that in the high-fugacity phase, the size of $Y^{1,2}$ are both $\sim\nu$ in the horizontal direction and ${\cal O}(1)$ in vertical.

Summarizing, we have shown that in holographic CFT a thermodynamic-like phase transition can arise from eigenstate thermalization, and falls into the Lee-Yang paradigm. It is possible to generalize our results to realistic quantum many-body systems, where CFT may cease to work. Indeed, let $H$ be a generic many-body Hamiltonian and $|E\rangle$ be its eigenstate. Consider the eigenstate auto-correlation $f_E(\tau)=\langle E|O(\tau)O(0)|E\rangle$ between operators $O$ at imaginary time $\tau$ and $0$. Thanks to $f_E(\tau)=e^{\tau E}\langle \psi_i|e^{-\tau H}|\psi_i\rangle$ with $|\psi_i\rangle=O(0)|E\rangle$, this observable essentially describes the amplitude of the overlap of the imaginary-time evolving wavefunction $e^{-\tau H}|\psi_i\rangle$ with its initial state $|\psi_i\rangle$. Strikingly, our result, when being generalizing here, becomes an imaginary-time version of the so-called dynamical quantum phase transition, that has received many theoretical and experimental attentions in recent years \cite{Heyl13,Heyl18}. In-depth explorations of this relation are desirable in the future.

We thank Tomoki Nosaka for collaborations at the initial stage of this project. We also thank Hongfei Shu, Futoshi Yagi, Ruidong Zhu for helpful discussions. This work is supported by National Science Foundation of China (NSFC) grants No.\,12175238 and No.\,12447108 (Y.X, W.S, H.W), No.\,12475043 and No.\,12447101 (C.T.).


\begin{thebibliography}{}
\bibitem{note_saddle} While the analysis of the saddle point equation are technically involved, we expose it as well as the explicit form of that equation in Ref.~\cite{Wang:2026}.
\end{thebibliography}

\end{document}